\documentclass[aps,prb,nofootinbib,superscriptaddress, floatfix,groupedaddress,longbibliography,preprint]{revtex4-1}
\usepackage[english]{babel}
\usepackage[utf8]{inputenc}
\usepackage{mathtools}
\usepackage{braket}
\usepackage{amsfonts}
\usepackage{lineno}
\usepackage{pdfpages}
\usepackage{stfloats}
\usepackage{xcolor}
\usepackage[symbol]{footmisc}
\makeatletter
\AtBeginDocument{\let\LS@rot\@undefined}
\makeatother

\usepackage[justification=raggedright,singlelinecheck=false]{caption,subcaption}
\usepackage{siunitx}

\usepackage{scalerel}
\usepackage{tikz}
\usetikzlibrary{svg.path}
\definecolor{orcidlogocol}{HTML}{A6CE39}
\tikzset{
  orcidlogo/.pic={
    \fill[orcidlogocol] svg{M256,128c0,70.7-57.3,128-128,128C57.3,256,0,198.7,0,128C0,57.3,57.3,0,128,0C198.7,0,256,57.3,256,128z};
    \fill[white] svg{M86.3,186.2H70.9V79.1h15.4v48.4V186.2z}
                 svg{M108.9,79.1h41.6c39.6,0,57,28.3,57,53.6c0,27.5-21.5,53.6-56.8,53.6h-41.8V79.1z M124.3,172.4h24.5c34.9,0,42.9-26.5,42.9-39.7c0-21.5-13.7-39.7-43.7-39.7h-23.7V172.4z}
                 svg{M88.7,56.8c0,5.5-4.5,10.1-10.1,10.1c-5.6,0-10.1-4.6-10.1-10.1c0-5.6,4.5-10.1,10.1-10.1C84.2,46.7,88.7,51.3,88.7,56.8z};}}
\newcommand\orcidicon[1]{\href{https://orcid.org/#1}{\mbox{\scalerel*{
\begin{tikzpicture}[yscale=-1,transform shape]
\pic{orcidlogo};
\end{tikzpicture}
}{|}}}}
\usepackage[colorlinks,citecolor=red,urlcolor=blue,bookmarks=false,hypertexnames=true]{hyperref} 
\usepackage[capitalise]{cleveref}

\newcommand{\kwv}{\mathbf{k}}

\newcommand{\vcm}{$\rm V \, cm^{-1}$}

\begin{document}

\title{Valleytronics and negative differential resistance in cubic boron nitride: a first-principles study}

\author{Benjamin Hatanp{\"a}{\"a} \orcidicon{0000-0002-8441-0183}}
\affiliation{Division of Engineering and Applied Science, California Institute of Technology, Pasadena, CA, USA}

\author{Austin J. Minnich \orcidicon{0000-0002-9671-9540}}
\thanks{Corresponding author: \href{mailto:aminnich@caltech.edu}{aminnich@caltech.edu}}

\affiliation{Division of Engineering and Applied Science, California Institute of Technology, Pasadena, CA, USA}

\date{\today} 

\begin{abstract}
Cubic boron nitride (c-BN) is an ultrawide-bandgap semiconductor of significant interest for high-frequency and high-power electronics applications owing to its high saturation drift velocity and high electric breakdown field. Beyond transistors, devices exploiting the valley degree of freedom or negative differential resistance are of keen interest. While diamond has been found to have potential for these applications, c-BN has not been considered owing to a lack of knowledge of the relevant charge transport properties.  Here, we report a study of the high-field transport and noise properties of c-BN using first-principles calculations. We find that c-BN exhibits an abrupt region of negative differential resistance (NDR) below 140 K, despite the lack of multi-valley band structure typically associated with NDR. This feature is found to arise from a strong energy dependence of the scattering rates associated with optical phonon emission. The high optical phonon energy also leads to an intervalley scattering time rivaling that of diamond. The negative differential resistance and long intervalley scattering time indicate the potential of c-BN for transferred-electron and valleytronic devices, respectively.

\end{abstract}

\maketitle

\noindent

\section{Introduction}






Ultrawide-bandgap (UWBG) semiconductors are the subject of intense study owing to their utility in power electronics and related device applications \cite{Tsao_2018,Xu_2022}. Cubic boron nitride (c-BN), with a wide bandgap of 6.4 eV \cite{Chrenko_1974}, has long been of interest for electron devices and other applications owing to its competitive mechanical, thermal, and electrical properties. c-BN has a high hardness of 30-43 GPa \cite{Harris_2004}, thermal conductivity which is second only to diamond \cite{Chen_2020}, excellent oxidation resistance, and high chemical and thermal stability. In addition, c-BN also has a high electric breakdown field of 4 MV $\text{cm}^{-1}$, comparable to those of diamond and GaN; \cite{Chillieri_2022_1} and a high predicted saturation drift velocity of 4.3 $\times \ 10^7$ cm $\text{s}^{-1}$, which is the highest of any semiconductor. These properties lead to the prediction of among the highest figure-of-merits (such as Johnson, Baliga, and Keyes figures-of-merit) for high-frequency and high-power applications \cite{Tsao_2018,Chillieri_2022_1,Haque_2024}.

In addition to transistor-based devices using UWBGs, other device types such as transferred-electron and valleytronic devices are of interest. For instance, diamond exhibits a region of negative differential resistance (NDR) below room temperature \cite{Isberg_2012}, enabling the realization of Gunn oscillators. \cite{Suntornwipat_2016} Further, the large intervalley time of 300 ns \cite{Isberg_2013} allows for manipulation of electrons by their valley degree of freedom, for instance in a valleytronic transistor \cite{Suntornwipat_2021}. However, diamond has long-standing challenges, including the difficulty to realize n-type doping \cite{Borst_1995,Gheeraert_2002,Shah_2008} and synthesis of high-quality thin films.

c-BN has potential to overcome some of these difficulties. c-BN can be doped both n- and p-type \cite{Taniguchi_2002,Wang_2003,Litvinov_1998,Taniguchi_2003}, has a higher oxidation temperature than diamond, and is more thermally and chemically stable \cite{Izyumskaya_2017}. However, c-BN films face various challenges in synthesis, including formation of nanocrystalline films, growth of the hexagonal crystal structure rather than the desired cubic structure \cite{Samantaray_2005}, and high compressive stresses. \cite{Yang_2011} As a result, devices based on c-BN are still rare. Most devices are limited to $p-n$ junctions \cite{Nose_2005} fabricated from doped or intrinsic c-BN thin films \cite{Zhang_2013}. Diodes \cite{Mishima_1987} and ultraviolet emitters \cite{Mishima_1988} have been realized in c-BN. In addition, deep-ultraviolet photodetectors based on c-BN have been fabricated \cite{Liao_2002,Soltani_2008,Benmoussa_2009}, which are of interest for extreme-environment applications.

Considering that diamond thin films have been found to be promising for Gunn oscillators and valleytronic transistors \cite{Suntornwipat_2016,Suntornwipat_2021}, it is natural to consider c-BN as well given the similarities in electronic band structure. However, experimental data regarding the transport properties of c-BN, especially at high electric fields, are scarce. The low-field mobility has been reported experimentally \cite{Mohammad_2002,Wang_2003,Hirama_2020,Haque_2021}, with values varying by several orders of magnitude. The low-field mobility of c-BN has been computed with ab-initio methods \cite{Sanders_2021,Ponce_2021,Iqbal_2024}, but investigations at higher fields are limited to Monte Carlo methods with semi-empirical inputs \cite{Siddiqua_2020,Chillieri_2022_low,Zhu_2023}.  The properties relevant for transferred-electron and valleytronic devices, namely the occurrence of negative differential resistance and a sufficiently long intervalley scattering time, have not yet been assessed in c-BN. 

Here, we report first-principles calculations of the high-field electron transport properties and noise characteristics of c-BN. We find a pronounced region of negative differential resistance below 140 K. This feature occurs due to the strong dependence of the electron scattering rates associated with optical phonon emission. The high optical phonon energy in c-BN also leads to intervalley scattering times rivaling those of diamond. We identify how these predictions could be experimentally tested via a non-monotonic trend in the spectral noise current density versus electric field. These properties suggest that c-BN is a promising contender in novel electronics applications such as valleytronics or Gunn oscillators.

\section{Theory and Numerical Methods}

The approach used here to compute the high-field transport and noise properties of electrons has been described previously \cite{Choi_2021,Cheng_2022,Hatanpaa_2023,Catherall_2023}. In brief, the Boltzmann equation for a non-degenerate and spatially homogeneous electron gas subject to an applied electric field is given by

\begin{equation}
    \frac{q\mathbf{E}}{\hbar} \cdot \nabla_{\mathbf{k}} f_{\mathbf{k}} = - \sum_{\mathbf{k}'} \Theta_{\mathbf{k}\mathbf{k}'} \Delta f_{\mathbf{k}'} 
\end{equation}

Here, $q$ is the carrier charge, $\mathbf{E}$ is the electric field vector, $f_{\mathbf{k}}$ is the distribution function describing the occupancy of the electronic state indexed by wavevector $\mathbf{k}$, $\Delta f_{\mathbf{k}'} $ is the perturbation to the equilibrium electron distribution function $f_{\mathbf{k}}^{0}$, and $\Theta_{\mathbf{k}\mathbf{k}'}$ is the linearized collision matrix arising from Fermi's golden rule given by Eq. 3 of Ref.~\cite{Choi_2021}. Here, we assume that only one band contributes to charge transport as the next-lowest conduction band minimum is calculated to be 4.32 eV higher in energy, far higher than the electronic states occupied at the highest fields used in this work, and thus we neglect the band index in our notation. The formulation of the BTE we use here is applicable to arbitrarily high fields \cite{Cheng_2022} so long as the electron gas remains nondegenerate.

Beyond the low-field regime, the reciprocal space derivative of the total distribution function must be evaluated numerically. Here, we use a finite difference approximation given in Refs.~\cite{Mostofi_2008, Marzari_2012}. The BTE then takes the form of a linear system of equations (Eqn.~5 in Ref.~\cite{Choi_2021}), which can be solved using numerical linear algebra:
\begin{equation}
    \sum_{\mathbf{k}'} \Lambda_{\mathbf{k}\mathbf{k}'} \Delta f_{\mathbf{k}'} = \sum_{\gamma} \frac{qE_{\gamma}}{k_{B}T} v_{\mathbf{k},\gamma} f_{\mathbf{k}}^{0} 
\end{equation}
Here, $E_{\gamma}$ and $v_{\mathbf{k},\gamma}$ are the electric field and electron drift velocity along the $\gamma$-Cartesian axis, and the relaxation operator $\Lambda_{\mathbf{k}\mathbf{k}'}$ is defined as 

\begin{equation}
    \Lambda_{\mathbf{k}\mathbf{k}'} =  \Theta_{\mathbf{k}\mathbf{k}'} + \sum_{\gamma} \frac{qE_{\gamma}}{\hbar} D_{\mathbf{k}\mathbf{k}',\gamma}
\end{equation}

where $D_{\mathbf{k}\mathbf{k}',\gamma}$ is the momentum-space derivative represented in the finite-difference matrix representation given by Eqn.~24 in Ref.~\cite{Mostofi_2008}. Once the steady-state distribution function is obtained by solving the linear system, various transport properties can be calculated via an appropriate Brillouin zone sum. For instance, the drift velocity in the $\beta$ direction is given by

\begin{equation}
    V_{\beta} = \frac{1}{N} \sum_{\mathbf{k}} v_{\mathbf{k},\beta} f_{\mathbf{k}} 
\end{equation}

where $N = \sum_{\mathbf{k}} f_{\mathbf{k}}$ is the number of electrons in the Brillouin zone. Similarly, the total intervalley scattering rate $\Xi_{\mathbf{k}}$ of a state $\mathbf{k}$ can be expressed as 
\begin{equation}
    \Xi_{\mathbf{k}} = \sum_{\mathbf{k}'} \Theta_{\mathbf{k}' \mathbf{k}} (1 - \delta_{\mathbf{k}' \mathbf{k}})
\end{equation}

Here, $\delta_{\mathbf{k}' \mathbf{k}} = 1$ if $\mathbf{k}'$ and $\mathbf{k}$ are in the same valley, and $\delta_{\mathbf{k}' \mathbf{k}} = 0$ if they are in different valleys. We do not make any distinction between valleys on the same axis as the minima in c-BN are at the $X$ point, meaning that such valleys are connected across the Brillouin zone edges. There thus exist three distinct valley types: (100), (010), and (001). We then define the average intervalley scattering time $\tau_{\text{int}}$ as 

\begin{equation}
    \tau_{\text{int}} = \frac{1}{N} \sum_{\mathbf{k}} \Xi_{\mathbf{k}} f_{\mathbf{k}}
\end{equation}

The current noise, characterized by the power spectral density (PSD), can also be computed using the BTE. As derived in Ref.~\cite{Choi_2021}, the current PSD $S_{j_{\alpha}j_{\beta}}$ can be calculated at angular frequency $\omega$ as

\begin{equation} \label{eq:Sj}
 S_{j_{\alpha}j_{\beta}}(\omega) = 2 \bigg( \frac{2 e}{\mathcal{V}_0}\bigg)^2 \Re \bigg[\sum_{\kwv} v_{\kwv,\alpha} \sum_{\kwv'} (i\omega \mathbb{I} + \Lambda)^{-1}_{\kwv \kwv'} \bigg(f_{\mathbf{k}'}^s (v_{\kwv',\,\beta} - V_{\beta})\bigg)\bigg]
\end{equation}
where $j_{\alpha}$ and $j_{\beta}$ are the current densities along axes $\alpha$ and $\beta$, and $\mathbb{I}$ is the identity matrix. In the limit $\omega \tau \ll 1$, where $\tau$ is a characteristic relaxation time, $S_{j_{\alpha}j_{\beta}}$ is proportional to the diffusion coefficient, a relation known as the fluctuation-diffusion relation \cite{GGK_1979}. We can thus compute the diffusion coefficient from \cref{eq:Sj}, and for simplicity we refer to the quantity defined in \cref{eq:Sj} as the diffusion coefficient.

The numerical details are as follows. For all calculations, the electron-phonon matrix elements and electronic structure are computed on a coarse $12 \times 12 \times 12$ grid using DFPT and DFT in \textsc{Quantum Espresso} \cite{Giannozzi_2009}. The PBE functional was used for the DFT calculations. A wave-function energy cutoff of 80 Ryd was used for all calculations. A relaxed lattice parameter of 3.623 $\textrm{\AA}$ was used, which overestimates the experimental value by only 0.22\% \cite{Knittle_1989}. The electronic structure and electron-phonon matrix elements were then interpolated to a fine grid using \textsc{Perturbo} \cite{Zhou_2021}. For all temperatures, a fine grid of $160 \times 160 \times 160$ was used with a 5 meV Gaussian smearing parameter. Increasing the grid size to $180 \times 180 \times 180$ led to a maximum change of  3.3\% in the mobility and maximum change of 8.0\% in the diffusion coefficient. An energy window of 383 meV was used. Increasing this energy window to 437 meV led to a maximum change of 0.16\% in the mobility and maximum change of 4.0\% in the diffusion coefficient. The linear system of equations used to compute the high-field transport properties was then solved by a Python implementation of the GMRES method \cite{Fraysse_2005}. For all calculations and temperatures, the Fermi level was adjusted to yield a carrier density of $4 \times 10^{13} \ \text{cm}^{-3}$. For all calculations of the diffusion coefficient, a frequency of 1 GHz was used, selected so to ensure that $\omega \tau^{-1} \ll 1$ (where $\tau$ is a characteristic relaxation time), while avoiding too low frequencies which result in numerical instabilities. 

\section{Results}

We begin by examining the dependence of the electron drift velocity and mobility on electric field at various temperatures. We first compare our computed low-field mobility with other reported values. At 300 K, we compute a low-field mobility of 1860 $\text{cm}^2 \text{V}^{-1} \text{s}^{-1}$. When including two-phonon scattering in the framework described in Refs.~\cite{Cheng_2022,Hatanpaa_2023}, the value decreases to 1136 $\text{cm}^2 \text{V}^{-1} \text{s}^{-1}$. c-BN has an experimentally reported Hall mobility value of 825 $\text{cm}^2 \text{V}^{-1} \text{s}^{-1}$ \cite{Wang_2003} (although much lower values have been reported \cite{Mohammad_2002,Hirama_2020,Haque_2021}), and previously computed ab-initio values range from 1230 $\text{cm}^2 \text{V}^{-1} \text{s}^{-1}$ (Ref. \cite{Iqbal_2024}) to 1610 $\text{cm}^2 \text{V}^{-1} \text{s}^{-1}$ (Ref. \cite{Sanders_2021}). Our mobility values are thus in reasonable agreement with prior computed and experimental values. 

We note that for the rest of this work, only one-phonon scattering is considered. We find that adding two-phonon scattering decreases the mobility at all temperatures by roughly 40\%, and the qualitative features of the mobility and diffusion coefficient versus electric field are generally retained. These findings are similar to those in Refs.~\cite{Cheng_2022,Hatanpaa_2023}. Therefore, to reduce computational cost, we consider only one-phonon scattering as is typically assumed.

\begin{figure}[h]
\includegraphics[]{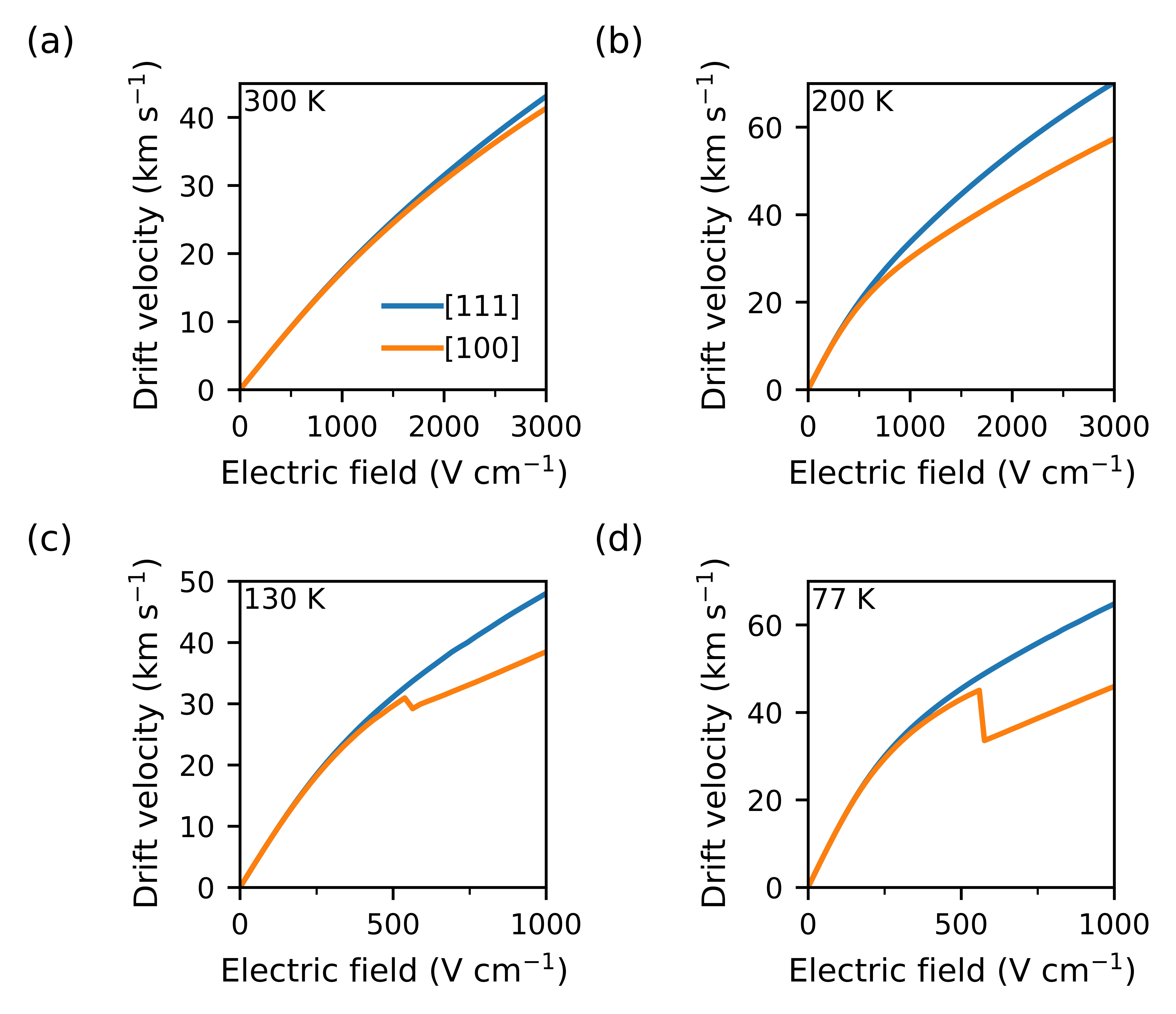}
{\phantomsubcaption\label{dv_vs_e_300}
\phantomsubcaption\label{dv_vs_e_200}
\phantomsubcaption\label{dv_vs_e_130}
\phantomsubcaption\label{dv_vs_e_77}}
\caption{Computed drift velocity versus electric field at (a) 300 K, (b) 200 K, (c) 130 K, and (d) 77 K, with field applied along the [100] direction (orange line) and [111] direction (blue line).}
\label{dv_vs_e}
\end{figure}

Next, we examine the drift velocity versus electric field at 300 K in \cref{dv_vs_e_300}. At sufficiently high electric fields, we observe an anisotropy in the drift velocity despite the cubic symmetry of the crystal, with the drift velocity in the [100] direction being less than that in the [111] by around 4\% at 3 kV $\text{cm}^{-1}$. This anisotropy is present with greater magnitude at 200 K, as seen in \cref{dv_vs_e_200}. We also note that as the temperature decreases, the onset of a discernible difference in drift velocity between the two directions occurs at a lower field. The anisotropy arises due to differences in the average energy of the electron distribution function between the longitudinal and transverse valleys \cite{conwell_1967,Canali_1975}, as discussed in more detail below.



At 130 K, a qualitative change in the [100] drift velocity characteristics is seen in \cref{dv_vs_e_130} as an abrupt decrease in drift velocity above $\approx 500$ V $\text{cm}^{-1}$. The effect is even more pronounced at 77 K, with the drift velocity dropping 25\% within only 20 V $\text{cm}^{-1}$. The decrease in drift velocity with increasing field is known as negative differential resistance (NDR). The effect is well-known in semiconductors such as GaAs, forming the basis for Gunn diodes which provide microwave power from a DC bias \cite{Ridley_1961,Gunn_1963}. For these materials, NDR is caused by intervalley scattering of electrons from the primary, high-mobility valley at $\Gamma$ to a satellite valley with higher effective mass such as the $L$ valley in GaAs. However, this explanation is not applicable in c-BN as we calculate that the next-lowest-energy satellite valley is 2.44 eV higher than the conduction band minimum. This energy is sufficiently large that it plays a negligible role in the transport in the electric field range used in this work.

In diamond, NDR has been experimentally observed when the electric field is oriented along the [100] direction, despite lacking the two-valley band structure required for the conventional NDR mechanism \cite{Isberg_2012}. In this case, NDR was attributed to the sudden onset of intervalley scattering associated with zone-edge longitudinal acoustic phonon emission, which causes an abrupt repopulation from valleys transverse to the electric field to those parallel to it\cite{Isberg_2012}. This repopulation occurs only at a high enough electric field value where the threshold for longitudinal acoustic intervalley phonon emission is reached. 

\begin{figure}[h]
\includegraphics[]{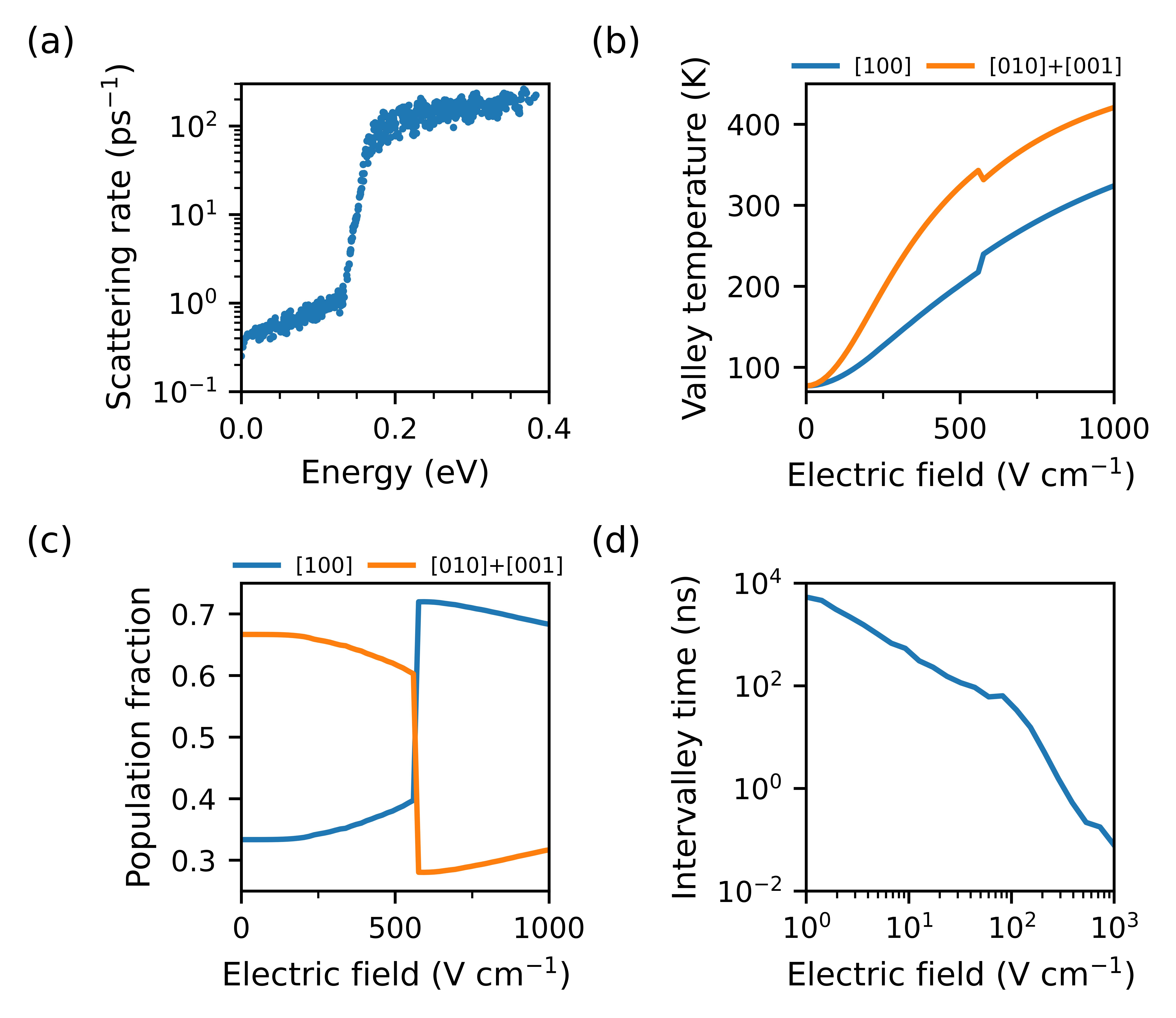}
{\phantomsubcaption\label{scattering_rates}
\phantomsubcaption\label{valley_temps}
\phantomsubcaption\label{population_frac}
\phantomsubcaption\label{intervalley_time}}
\caption{(a) Electron scattering rate versus energy. (b) Valley temperature versus electric field for [100] type valleys (blue), and [010]+[001] type valleys (orange). (c) Population fraction versus electric field for c-BN for [100] type valleys (blue) and the sum of the [010] and [001] type valleys (orange). (d) Average intervalley relaxation time versus electric field. All calculations are at 77 K with the electric field applied in the [100] direction. The intervalley time is on the order of microseconds at low fields, rivaling that of diamond.}
\label{fig2}
\end{figure}

As c-BN has a similar band structure to diamond with six equivalent conduction band minima, we hypothesize that a similar explanation is applicable. To test the hypothesis, we computed the energy dependence of the electron-phonon scattering rates. The result at 77 K is shown in \cref{scattering_rates}. Indeed, we observe that below the optical phonon energy of 150 meV, the scattering rates have a relatively weak dependence on energy, while the scattering rates abruptly increase above the optical phonon energy ($\sim 150$ meV). 

The strong energy-dependence of the scattering rates has consequences for the relative occupation in the various valleys and ultimately the transport properties. Below the threshold field, few electrons have sufficient energy to scatter to an inequivalent valley, and the [100], [010], and [001] valleys react to the electric field largely independently. In addition, owing to the lower effective mass of the transverse valleys relative to longitudinal ones (0.36$m_0$ versus 0.95$m_0$ for transverse and longitudinal, respectively), transverse valleys have a higher effective carrier temperature than the longitudinal valleys at a given field, as shown in \cref{valley_temps}. The effective valley temperatures are defined by computing the average energies of the steady distributions at the given field and identifying the temperature of a Boltzmann distribution with the same average energy. The valley temperature is a measure of the average energy of the distribution rather than a thermodynamic temperature. As the field increases, the transverse valleys achieve a higher steady-state temperature owing to their lower effective mass. The higher electron temperature of the transverse valleys also leads to the drift velocity anisotropy between [100] and [111] directions shown in \cref{dv_vs_e}.



At the threshold field, electrons in the transverse valleys gain enough energy to emit optical phonons and undergo an intervalley transition to the longitudinal valleys. This repopulation effect is shown in \cref{population_frac} at 77 K, which plots the population of the [100] type valleys in comparison to the sum of the [010] and [001] type valleys versus electric field, with the field applied in the [100] direction. At zero field, all valleys have the same population. As the field increases from zero, there is little population redistribution due to the absence of a significant intervalley scattering by phonon emission and the weak absorption-mediated scattering at 77 K. Even at 500 V $\text{cm}^{-1}$, the [100] type valleys have less than 40\% of the population. However, once the threshold field is reached ($\approx 560$ V $\text{cm}^{-1}$), a redistribution to the [100] longitudinal valleys occurs. Because the longitudinal mass is higher than the transverse mass, the drift velocity abruptly decreases. This repopulation effect occurs at all temperatures when the field is applied in any cubic axis, but NDR only manifests in the transport properties when phonon-absorption intervalley scattering is negligible compared to emission-mediated scattering, in this case at temperatures below around 140 K.


The high optical phonon energy in c-BN has consequences for intervalley scattering, which must be mediated by zone-edge modes due to momentum conservation. For sufficiently low fields, most electrons  do not have sufficient energy to scatter via an intervalley phonon emission process. Further, the high optical phonon energy of c-BN leads to low thermal occupation even at 300 K, relative to conventional semiconductors like GaAs. As a result, both absorption and emission-mediated intervalley scattering are weak in c-BN, and so the corresponding average intervalley relaxation time is expected to be long in comparison to other semiconductors with lower phonon energies. 

In \cref{intervalley_time}, we show the intervalley time versus electric field along the [100] direction at 77 K. At low field ($\approx$1 V $\text{cm}^{-1}$), the intervalley relaxation time is calculated to be 5.3 $\mu$s. For comparison, we also computed the intervalley relaxation time in diamond, obtaining a value of 2.4 $\mu$s. c-BN thus has an intervalley relaxation time at low fields nearly 100\% larger than that of diamond. We note that the value decreases with increasing electric field as electrons are able to emit zone-edge phonons and scatter to other valleys. However, the relatively long intervalley time suggests that c-BN may be promising in valleytronic applications.

Experimental tests of these predictions are challenging owing to the difficulties in preparing high-quality thin films of c-BN. We suggest an approach to mitigate this challenge based on measurement of the current noise power spectral density (PSD), or equivalently in the low-frequency limit, the diffusion coefficient. Due to the long intervalley scattering time in c-BN, we expect a clear intervalley noise contribution arising from electrons scattering between valleys with distinct effective masses when the electric field is applied in the [100] direction. Intervalley noise manifests in experiment as an anisotropy in the diffusion coefficient, with the value being larger along the direction with inequivalent valleys relative to the case in which all valleys are equivalent \cite{Price_1960, Hartnagel_2001}([100] valleys versus [111] valleys, respectively, for c-BN). 

More precisely, the general expression for the intervalley diffusion coefficient is given by 

\begin{equation} \label{eq:int}
D^{\text{int}} = n_1 n_2 (v_1 - v_2)^2  \tau_{\text{int}}
\end{equation}

 
Here, $n_1$ and $n_2$ are the fractions of electrons in valleys of type 1 and 2, $v_1$ and $v_2$ are the drift velocities in valleys of type 1 and 2, and $\tau_i$ is the intervalley relaxation time \cite{Brunetti_1981}. In the present case, the valley types refer to longitudinal and transverse valleys defined by the field direction. Thus, intervalley noise will manifest as an increase in the [100] diffusion coefficient over the [111] diffusion coefficient, with the precise amount depending on the quantities in \cref{eq:int} at each field.

\begin{figure}[h]
\includegraphics[]{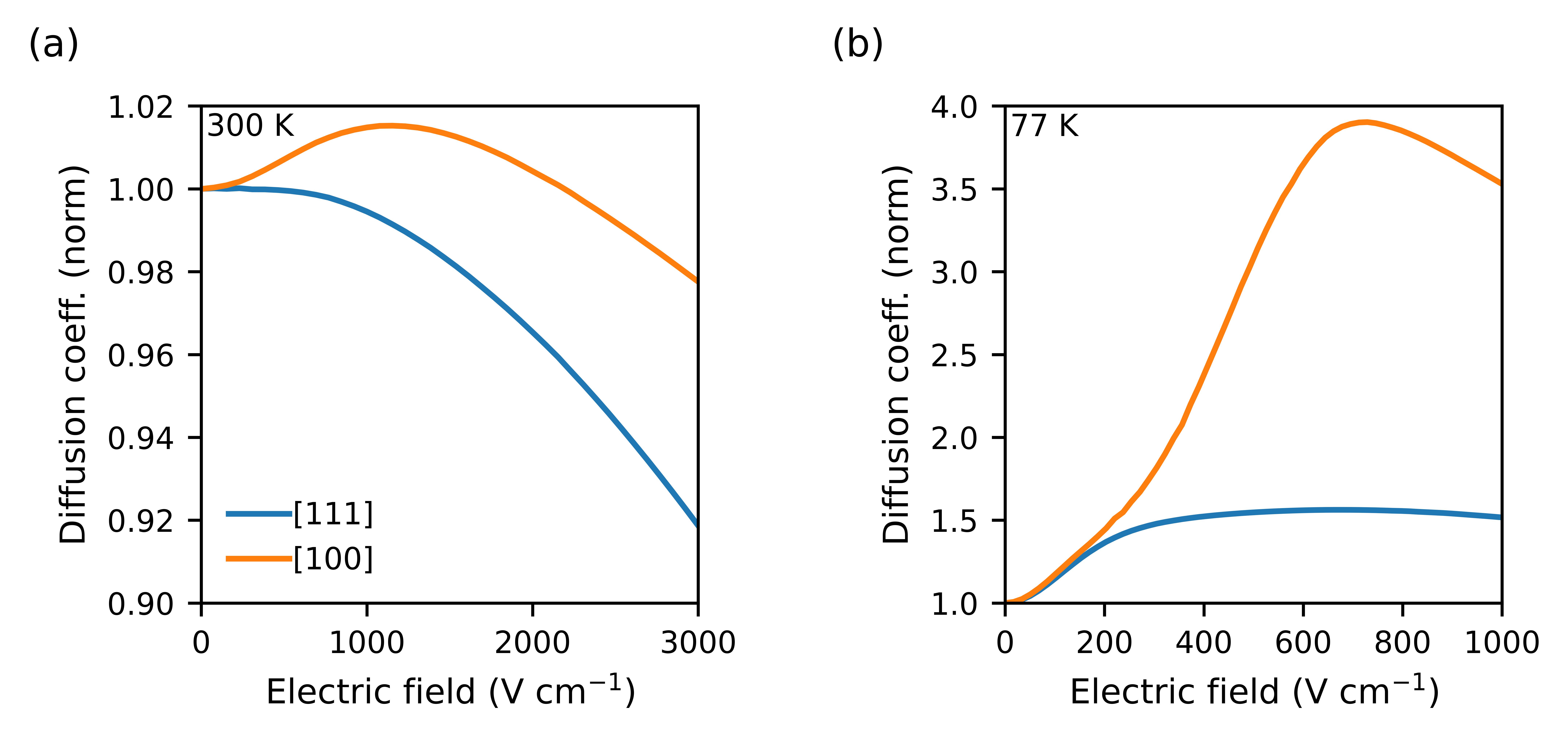}
{\phantomsubcaption\label{psd_vs_e_300}
\phantomsubcaption\label{psd_vs_e_77}}
\caption{Computed electron diffusion coefficient versus electric field at (a) 300 K and (b) 77 K, with field applied along the [100] direction (orange line) and [111] direction (blue line). A pronounced peak in the [100] diffusion coefficient is observed at 77 K, which is attributed to intervalley diffusion.}
\label{fig3}
\end{figure}

In \cref{fig3}, we show the electric field dependence of the diffusion coefficient for 300 K and 77 K. At 300 K in \cref{psd_vs_e_300}, we observe that the diffusion coefficient monotonically decreases with increasing field along the [111]. This behavior is similar to what is seen in n-Si \cite{Hatanpaa_2024}, and it occurs when the scattering rates increase sufficiently strongly with increasing energy \cite{Brunetti_1981,Aninkevicius_1993,Choi_2021}. For the [100] direction, however, a higher diffusion coefficient value and a non-monotonic trend with electric field are seen. At 300 K in \cref{psd_vs_e_300}, we observe that the diffusion coefficient peak when the electric field is applied in the [100] direction leads to a non-monotonic trend, with a peak around 2\% higher than the equilibrium value. We attribute this peak to the contribution of intervalley noise, although the magnitude of the peak may be difficult to detect experimentally.

\Cref{psd_vs_e_77} shows the corresponding results at 77 K. Along the [111], a non-monotonic trend is observed despite the expected absence of intervalley noise. This feature can be explained using an approximate generalization of the Einstein relation applied to high fields, $D = (2/3) \mu(E) (\langle \epsilon \rangle/e)$, where $\mu(E)$ is the electric field-dependent mobility and $\langle \epsilon \rangle$ the average electron energy \cite{Jacoboni_2010}. Due to weak dependence of scattering rates on energy below the optical phonon energy as seen in \cref{scattering_rates}, the average electron energy exhibits a stronger dependence on electric field than the mobility. As a result, the diffusion coefficient initially increases with increasing field. 

In the [100] direction, the peak is markedly larger compared to the [111] case, with the peak value of the [100] diffusion coefficient at $\approx$700 \vcm \ being nearly 300\% larger than the equilibrium value. The effect is much larger at 77 K compared to at 300 K. Such large peaks in the diffusion coefficient have been observed experimentally for other materials such as GaAs \cite{ruchkino1968,Bareikis_1980,gasquet1985} but have not been seen in first-principles calculations \cite{Cheng_2022} until now. Monte Carlo calculations of the diffusion coefficient in diamond at 300 K have found a slight increase of the diffusion coefficient with electric field \cite{Osman_1991}, similar to what is observed here in c-BN at 300 K. The magnitude of the peak at low temperatures is sufficiently large that it could easily be discerned in experiment, and its detection would support the prediction of the long intervalley time in c-BN.

\section{Discussion}

Our first-principles calculations have predicted that c-BN exhibits a region of NDR and a long intervalley lifetime rivaling that of diamond. These properties may find useful device applications. In materials with a pronounced NDR region, instabilities in electric current will lead to the formation of charged domains \cite{Gunn_1963}, which can be utilized in Gunn oscillators for various microwave applications. As a Gunn oscillator has been constructed with diamond thin films \cite{Suntornwipat_2016} it is possible that such devices could be realized in c-BN. We note that in both cases, the devices would need to operate below room temperature.


Experimental characterization of the NDR region can also provide insight into the role of 2ph scattering in c-BN.  While in this work we have employed the 1ph level of theory, 2ph scattering was found to reduce the predicted mobility by $\approx 40$\%, in line with the reduction reported for other semiconductors \cite{Lee_2020,Hatanpaa_2023,Esho_2023}. 2ph scattering also shifts the electron distribution to lower energies at a given field, which in turn would cause the threshold field for NDR to occur at $\approx 1120$ V $\text{cm}^{-1}$ than the 560 V $\text{cm}^{-1}$ at the 1ph level of theory. If sufficiently pure samples were available, the difference in threshold field should be discernible. 



For valleytronics, it is essential that electrons within a valley remain there long enough to perform the desired function \cite{Vitale_2018}. For c-BN, at equilibrium, our calculated intervalley time of 5.3 $\mu$s at 77 K is significantly greater than the 300 ns value for diamond, calculated from Monte Carlo simulation \cite{Isberg_2013} and the value of 2.4 $\mu$s we computed. It also greatly exceeds the intervalley time of typical semiconductors; for instance, we compute the intervalley time of n-Si to only be 149 ps at the same temperature. However, due to the strong decrease of the intervalley time with increasing field seen in \cref{intervalley_time}, some optimization may be required to use c-BN in valleytronic applications. A potential route to increase the intervalley relaxation time further is by leveraging the compressive strain present in most c-BN thin films. Strain would break the degeneracy of the six equivalent valleys, further inhibiting intervalley scattering and leading to an increase in intervalley relaxation time.



\noindent
\section{Summary} 

We have computed high-field transport properties and diffusion coefficient of c-BN from first principles from 77 -- 300 K. We find that below 140 K, c-BN exhibits a region of negative differential resistance arising from the strong energy dependence of the scattering rates around the optical phonon energy. The calculated intervalley time is comparable to that of diamond, suggesting that c-BN could be a promising material for valleytronic applications. We also show that our predictions can be tested by identifying a non-monotonic trend of the diffusion coefficient versus electric field. Our work highlights the potential electron device applications of c-BN beyond conventional power electronics, stimulating further experimental investigation into the synthesis and electrical transport properties of c-BN thin films.


\begin{acknowledgements}
B.H. was supported by a NASA Space Technology Graduate Research Opportunity. A.J.M. was supported by AFOSR under Grant Number FA9550-19-1-0321.
\end{acknowledgements}

\bibliography{references}

\end{document}